\documentclass[a4paper,twoside,english,superscriptaddress,showpacs,preprint]{revtex4}
\usepackage{mathptmx}

\usepackage[T1]{fontenc}
\usepackage[latin9]{inputenc}
\usepackage{color}
\usepackage{amsmath}
\usepackage{graphicx}
\usepackage{amssymb}

\makeatletter

\newcommand{\lyxdot}{.}

\@ifundefined{textcolor}{}
{%
 \definecolor{BLACK}{gray}{0}
 \definecolor{WHITE}{gray}{1}
 \definecolor{RED}{rgb}{1,0,0}
 \definecolor{GREEN}{rgb}{0,1,0}
 \definecolor{BLUE}{rgb}{0,0,1}
 \definecolor{CYAN}{cmyk}{1,0,0,0}
 \definecolor{MAGENTA}{cmyk}{0,1,0,0}
 \definecolor{YELLOW}{cmyk}{0,0,1,0}
 }

\makeatother

\usepackage{babel}

\begin{document}

\title{Material-barrier Tunneling in One-dimensional Few-boson Mixtures}

\author{Anika C. Pflanzer}

\affiliation{Physikalisches Institut, Universität Heidelberg, Philosophenweg 12,
69120 Heidelberg, Germany}

\author{Sascha Zöllner}

\email{zoellner@nbi.dk}

\affiliation{Niels Bohr International Academy, Niels Bohr Institute, Blegdamsvej
17, 2100 København, Denmark}

\author{Peter Schmelcher}

\email{peter.schmelcher@pci.uni-heidelberg.de}

\affiliation{Physikalisches Institut, Universität Heidelberg, Philosophenweg 12,
69120 Heidelberg, Germany}

\affiliation{Theoretische Chemie, Universität Heidelberg, Im Neuenheimer Feld
229, 69120 Heidelberg, Germany}
\begin{abstract}
We study the quantum dynamics of strongly interacting few-boson mixtures
in one-dimensional traps. If one species is strongly localized compared
to the other (e.g., much heavier), it can serve as an effective potential
barrier for that mobile component. Near the limit of infinite localization,
we map this to a system of identical bosons in a double well. For
realistic localization, the backaction of the light species on the
{}``barrier'' atoms is explained---to lowest order---in terms of
an induced attraction between these. Even in equilibrium, this  may
outweigh the bare intra-species interaction, leading to unexpected
correlated states. Remarkably, the backaction drastically affects
the inter-species dynamics, such as the tunneling of an attractively
bound pair of fermionized atoms.

\end{abstract}

\date{September 9, 2009}


\maketitle

Tunneling of particles through an energetically forbidden region is
a hallmark quantum-mechanical effect, which illustrates the wave nature
of matter. The experimental flexibility of ultracold atoms makes it
possible to study this phenomenon in a clean and highly controlled
environment, where the role of the tunnel barrier is played, e.g.,
by light forces such as in optical lattices \citep{bloch07}. This
has fostered the direct observation of fundamental effects like second-order
tunneling \citep{foelling07}, Josephson oscillations and nonlinear
self-trapping \citep{albiez05}. Moreover, optical lattices have
proven powerful quantum simulators, giving insight into the role of
tunneling in, e.g., the quantum phase transition from superfluid to
insulator \citep{greiner02}, or in spin-exchange processes responsible
for quantum magnetism \citep{trotzky08}. 

All of these cases reflect the paradigm of tunneling through a \emph{classical}
barrier described by some external potential. In this Letter, we investigate
tunneling through a material or \emph{quantum} barrier, in the sense
that it interacts with the particles. This is realized via quasi--one-dimensional
(1D) mixtures of two atomic species, one of which is squeezed in the
trap center. When the localization is very tight, we show that this
can indeed be understood as an effective tunnel barrier for the other
species. As that confinement is relaxed, the {}``barrier'' atoms
move due to the backaction of the other species. The dramatic effect
of this correlation on the mobile species is studied both for the
ground state and for the inter-species tunneling dynamics. To first
explore its microscopic mechanism, we consider a few-atom system system,
which is studied in a numerically exact fashion utilizing the multi-configuration
time-dependent Hartree method \citep{mey90:73}. This wave-packet
dynamics tool which has been applied successfully to systems of identical
bosons as well as mixtures (see \citep{zoellner08b} for details).
We then go on to derive approximate models that capture the relevant
physics for higher atom numbers. Still, such a small system may be
achieved experimentally, e.g., by creating arrays of 1D optical lattices
(such as in \citep{trotzky08}) or of 1D tubes \citep{haller09},
each containing only a few atoms. Moreover, high-resolution imaging
techniques such as scanning-electron microscopy allow for single-site
addressability \citep{gericke08}.

\paragraph*{Model.---}

We consider a mixture of two 1D bosonic species, labeled $\sigma=A,B$.
These may correspond to different atomic species (or isotopes); however,
in the case of equal masses they can also be thought of as different
hyperfine components. The many-body Hamiltonian then reads $H=\sum_{\sigma}H_{\sigma}+H_{\mathrm{AB}}$,
with the single-species Hamiltonian $H_{\sigma}=\sum_{i=1}^{N_{\sigma}}\left[\frac{1}{2m_{\sigma}}p_{\sigma,i}^{2}+U_{\sigma}(x_{\sigma,i})\right]+\sum_{i<j}g_{\sigma}\delta(x_{\sigma,i}-x_{\sigma,j})$
and the inter-species coupling $H_{\mathrm{AB}}=\sum_{i,j}g_{\mathrm{AB}}\delta(x_{\mathrm{A},i}-x_{\mathrm{B},j})$.
In what follows, we will focus on the case of harmonic trapping potentials
$U_{\sigma}(x)=\frac{1}{2}m_{\sigma}\omega_{\sigma}^{2}x^{2}$ and
repulsive forces, $g_{\sigma},g_{\mathrm{AB}}\ge0$. (Note that taking
$g_{\sigma}\to\infty$, the component $\sigma$ can be mapped to a
fermionic one \citep{girardeau60}, referred to as \emph{fermionization}.
In this sense, considering bosonic atoms poses no restriction in 1D.)
Rescaling to harmonic-oscillator units, we can eliminate $\hbar=m_{\mathrm{A}}=\omega_{\mathrm{A}}=1$
\citep{zoellner08b}.

\paragraph*{Material tunneling barrier.---}

Imagine the situation where one of the species (say B) is much more
strongly localized in the trap center. This can be achieved by drastically
reducing its (unperturbed) length scale $a_{\mathrm{B}}\equiv1/\sqrt{m_{\mathrm{B}}\omega_{\mathrm{B}}}\ll1$,
which amounts to having a near-zero mass ratio $\alpha\equiv m_{\mathrm{A}}/m_{\mathrm{B}}\ll1$
(for different atom masses such as in Li/Cs) and/or strong confinement
of B, $\omega_{\mathrm{A}}/\omega_{\mathrm{B}}\ll1$ (e.g., realized
via species-dependent optical lattices \citep{mandel03}). For simplicity,
let us assume much heavier B atoms but equal frequencies. This is
not crucial, though, and experimentally a large frequency $\omega_{B}$
may enhance the localization effect. 

Being tightly localized, the B atoms should feel no density variations
of A and, to lowest order, the total density matrix $\hat{\rho}_{\mathrm{AB}}^{(N)}\approx\hat{\rho}_{\mathrm{A}}^{(N_{\mathrm{A}})}\otimes\hat{\rho}_{\mathrm{B}}^{(N_{\mathrm{B}})}$
can be approximately factorized. (This even holds for strong repulsion
$g_{\mathrm{AB}}\gg1$, since the strong localization of the B atoms
at the trap center effectively reduces the interaction to a single-particle
potential \citep{pflanzer09a}.) Integrating out the heavy B atoms
leads to an effective Hamiltonian for the light species \citep{zoellner08b},
\begin{equation}
\!\bar{H}_{A}^{(0)}\negmedspace=\! H_{\mathrm{A}}+\mathrm{tr_{B}}[H_{\mathrm{AB}}\hat{\rho}_{\mathrm{B}}^{(N_{\mathrm{B}})}]\!=\! H_{\mathrm{A}}+g_{\mathrm{AB}}\!\sum_{i}\! n_{\mathrm{B}}(x_{\mathrm{A},i}).\negthickspace\negthickspace\label{eq:H_A}\end{equation}
In this light, A ought to feel only an effective single-particle potential
$g_{\mathrm{AB}}$$n_{\mathrm{B}}(x)$, which in our case likens a
sharp barrier at $x=0$, with a width $\sim a_{\mathrm{B}}$ given
by the one-body density $\rho_{\mathrm{B}}\equiv n_{\mathrm{B}}/N_{\mathrm{\mathrm{B}}}$
(in terms of the number density $n_{\mathrm{B}}$) and its height
proportional to the inter-species coupling $g_{\mathrm{AB}}$. In
particular, for $N_{\mathrm{B}}g_{\mathrm{AB}}\gg1$, this barrier
practically splits the harmonic trap for the A atoms into a pronounced
double well. (By extension, localizing the B atoms over several centers
should make for an effective {}``lattice'' for A.) This scenario
naturally brings up the question whether one can create an effective
localization of the light atoms (in equilibrium) or even see dynamic
effects such as tunneling solely due to inter-species interactions.

\begin{figure}
\includegraphics[width=0.33\columnwidth,keepaspectratio]{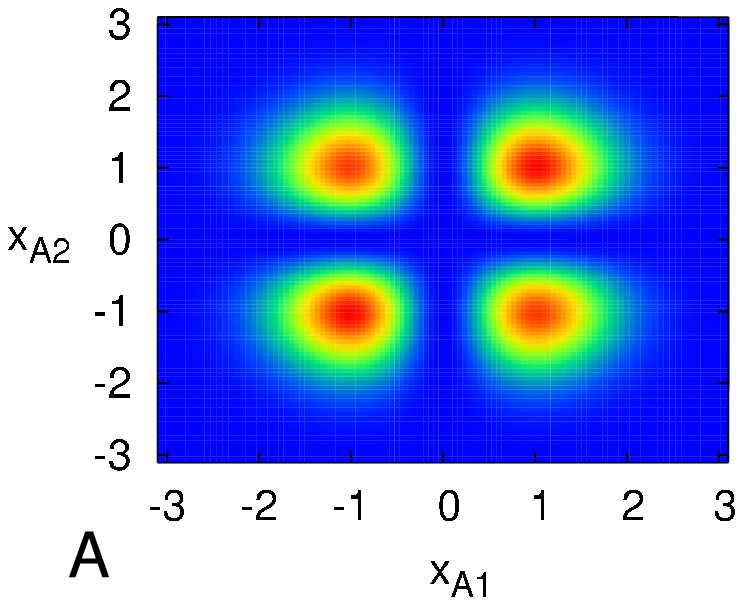}\includegraphics[width=0.33\columnwidth,keepaspectratio]{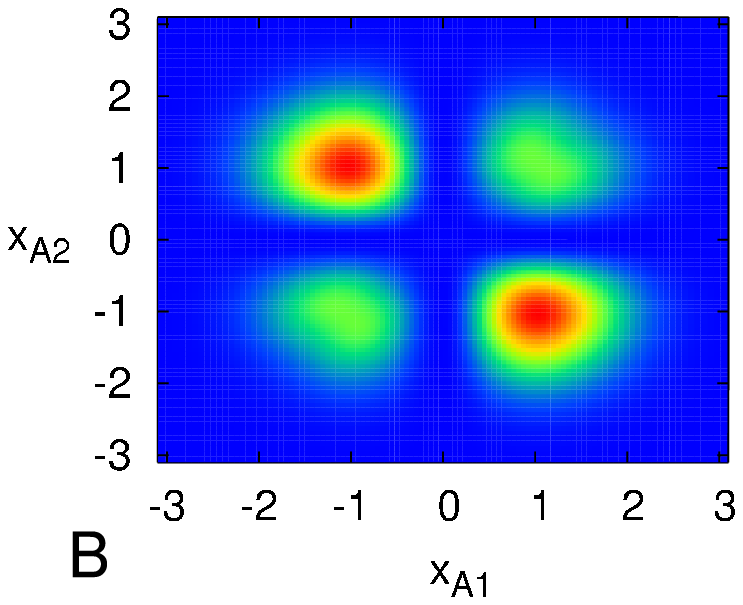}\includegraphics[width=0.33\columnwidth,keepaspectratio]{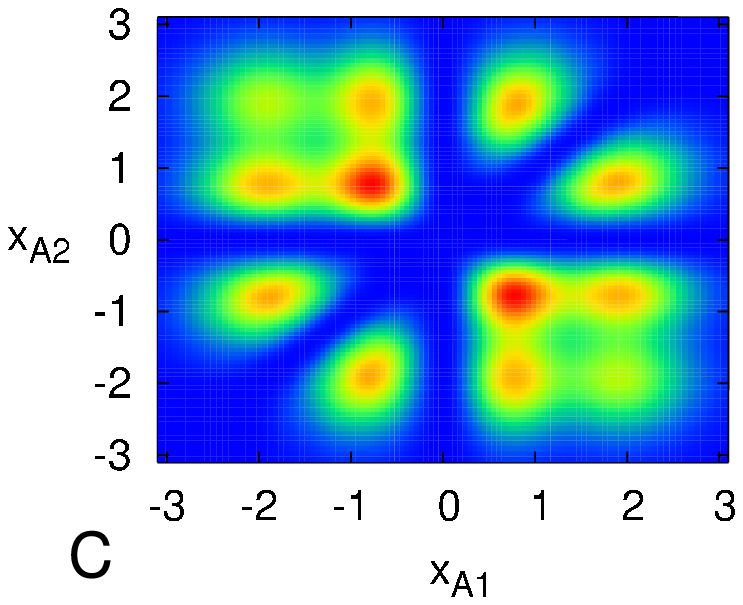}

\noindent \begin{centering}
\includegraphics[width=0.8\columnwidth,keepaspectratio]{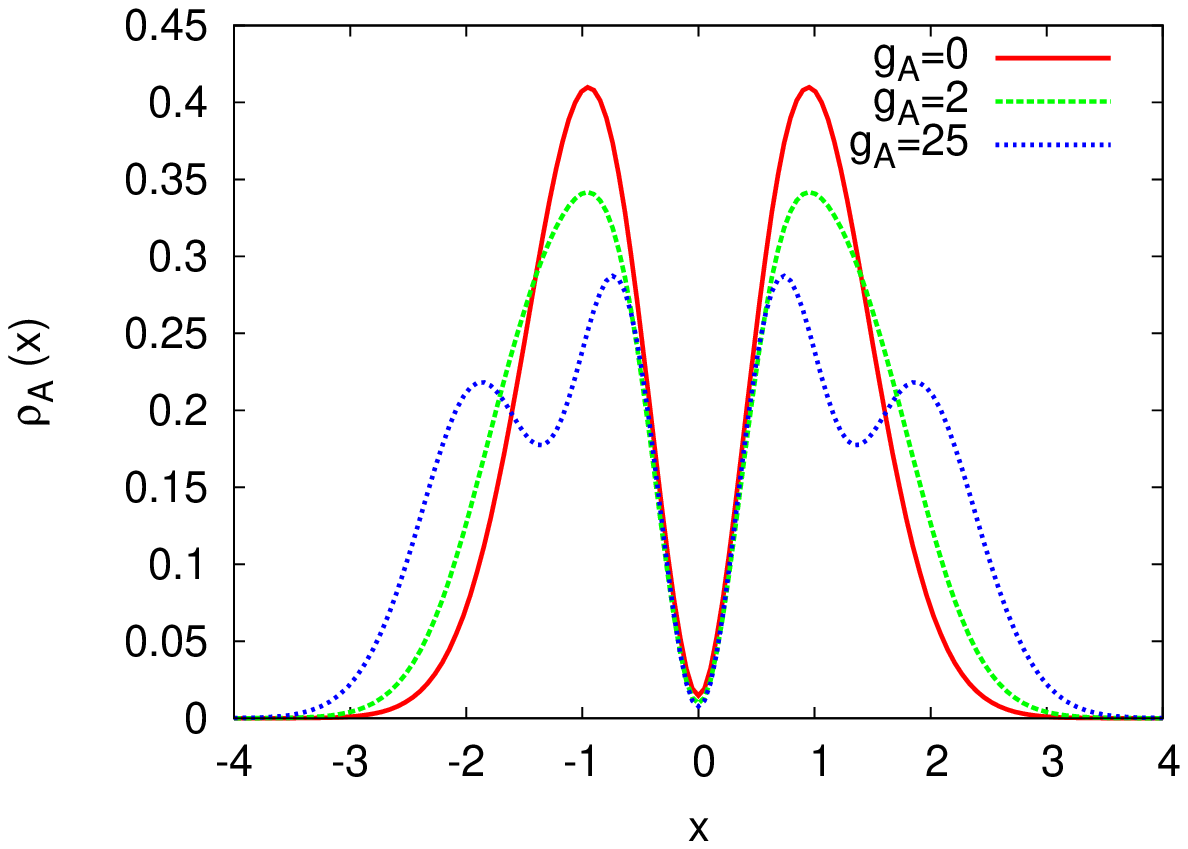}
\par\end{centering}

\caption{(color online) Strong-localization limit, $\alpha\equiv m_{\mathrm{A}}/m_{\mathrm{B}}=0.001$,
for $N_{\mathrm{A}}=4$ atoms in the presence of a B atom localized
at the center. \emph{Top}: Two-body density $\rho_{\mathrm{A}}^{(2)}(x_{1},x_{2})$
for $g_{\mathrm{A}}=0,\,0.5$, and $25$ from left to right. \emph{Bottom}:
Density profile $\rho_{\mathrm{A}}(x)$ for different $g_{\mathrm{A}}$.
(All quantities in harmonic-oscillator units of A, see text.) \label{cap:inf-mass}}

\end{figure}

Figure~\ref{cap:inf-mass} shows that, near the limit of strong localization,
the simple picture invoked above is indeed consistent with our exact
results. In detail, we have evidenced this on the ground state of
$N_{\mathrm{A}}=4$ atoms in the presence of one localized {}``impurity''
($\alpha=0.001$). Absent any intra-species interactions ($g_{\mathrm{A}}=0$),
the A atoms are coherently spread over the left- and right-hand side
of the trap in analogy to an external double-well potential \citep{zoellner06a},
even though the system as a whole is strongly interacting ($g_{AB}=25$
unless otherwise stated). This materializes in the two-body density
$\rho_{\mathrm{A}}^{(2)}(x_{1},x_{2})$, measuring the joint probability
density of finding two A atoms at $x_{1}$ and $x_{2}$: In Fig.~\ref{cap:inf-mass}(a),
both are equally likely to be on the same site (peaks on the diagonal,
$x_{1}\approx x_{2}$) as on opposite ones ($x_{1}\approx-x_{2}$).
As $g_{\mathrm{A}}$ increases, the A atoms assume a {}``Mott-insulator''-type
state with all A atoms more or less localized in either well {[}$g_{\mathrm{A}}=0.5$;
Fig.~\ref{cap:inf-mass}(b){]}, and eventually \emph{fermionize}
{[}$g_{\mathrm{A}}=25$; Fig.~\ref{cap:inf-mass}(c){]}. The checkerboard
pattern emerging in the latter case can be understood simply as a
gas of A-component bosons with hard-core repulsion (or, equivalently,
noninteracting fermions) immersed in a \emph{double-well} trap \citep{zoellner06a,goold08}:
In each well, measuring one A-boson at $x_{1}$ pins down the position
of the remaining bosons to $N_{\mathrm{A}}-1$ discrete spots. This
reflects in the density profile $\rho_{\mathrm{A}}(x)$, where $N_{\mathrm{A}}$
marked density maxima form as $g_{\mathrm{A}}\to\infty$ (Fig.~\ref{cap:inf-mass},
bottom).

\paragraph*{Beyond the static limit.---}

\begin{figure}
\includegraphics[width=0.33\columnwidth,keepaspectratio]{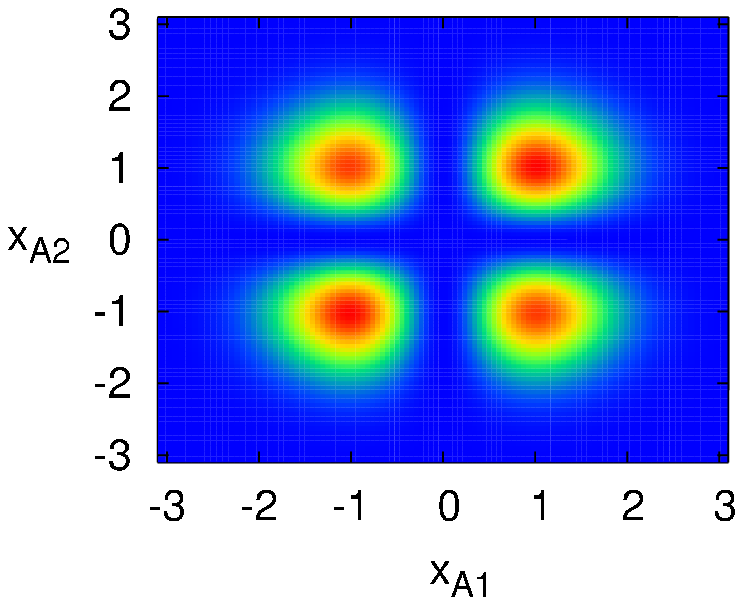}\includegraphics[width=0.33\columnwidth,keepaspectratio]{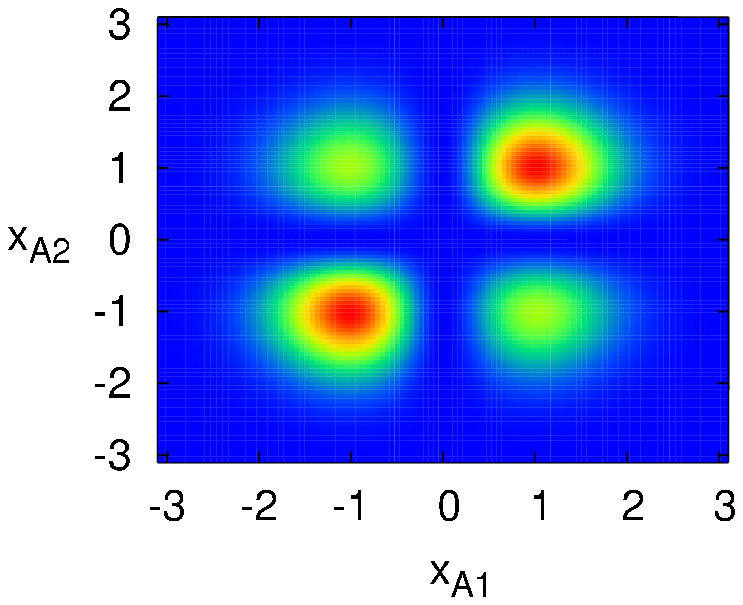}\includegraphics[width=0.33\columnwidth,keepaspectratio]{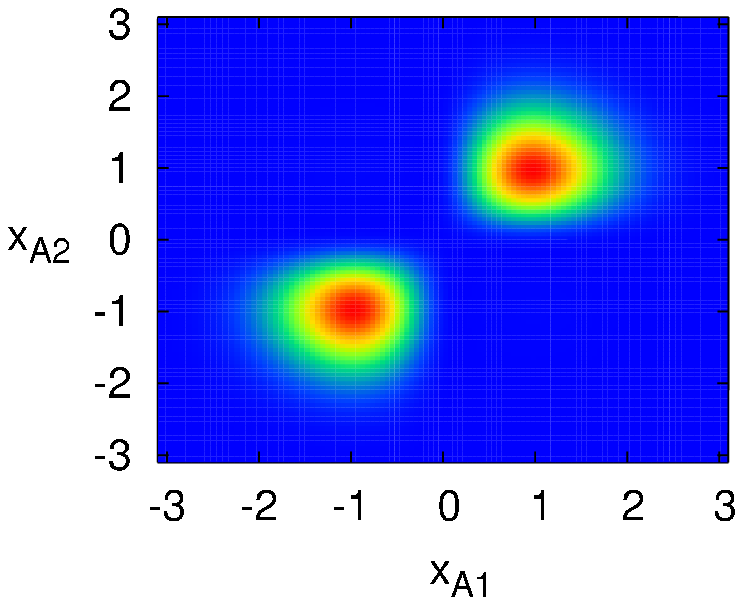}

\includegraphics[width=0.33\columnwidth,keepaspectratio]{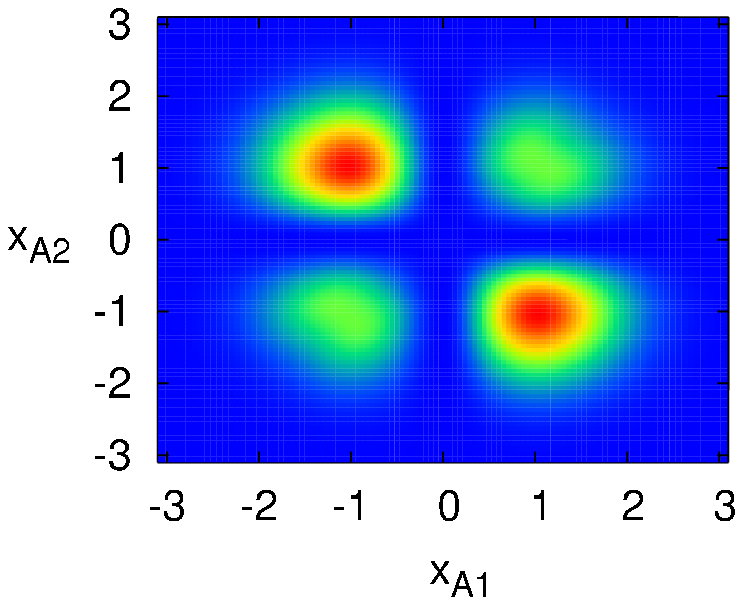}\includegraphics[width=0.33\columnwidth,keepaspectratio]{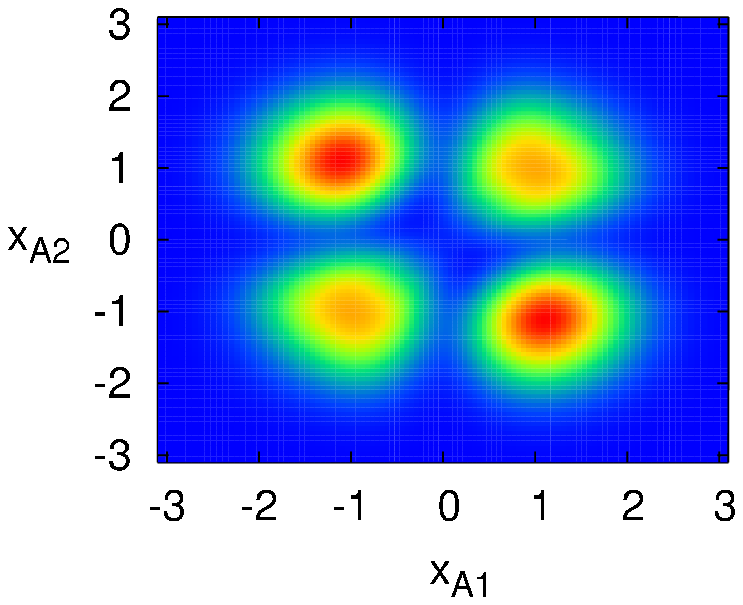}\includegraphics[width=0.33\columnwidth,keepaspectratio]{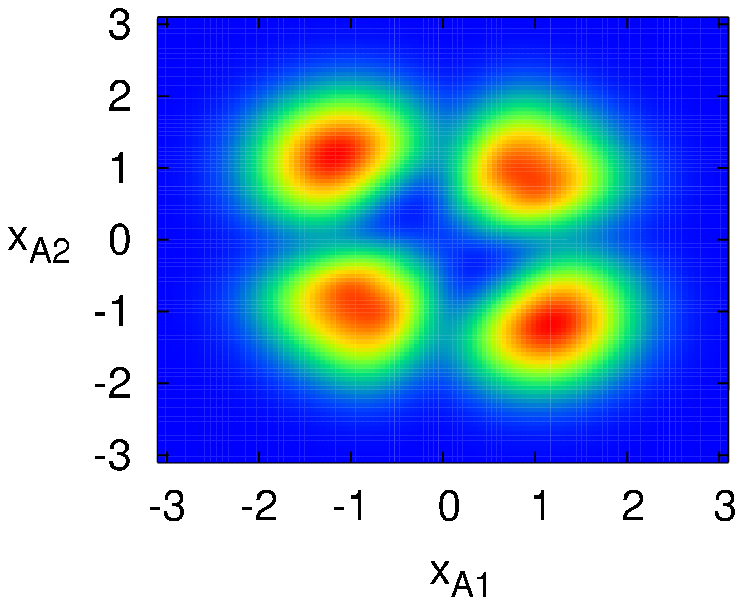}

\caption{(color online) Beyond the static-barrier picture: Two-body density
$\rho_{\mathrm{A}}^{(2)}(x_{1},x_{2})$ for (\emph{top}) $g_{\mathrm{A}}=0$,
with increasing mass ratios $\alpha=0.001,\,0.002,\,0.02$ (\emph{from
left to right}); (\emph{bottom}) $g_{\mathrm{A}}=0.5$, $\alpha=0.001,\,0.12,\,0.2$.
Same atom numbers as in Fig.~\ref{cap:inf-mass}. \label{cap:GS-alpha}}

\end{figure}

We have so far proceeded on the assumption that the B atoms were heavy
enough to be frozen out completely, that is, treated as a classical
potential. For less restrictive mass ratios, however, these are expected
to move due to the backaction of the light atoms. The impact of this
correlation becomes palpable in Fig.~\ref{cap:GS-alpha}, where the
ground-state evolution of $\rho_{A}^{(2)}(x_{1},x_{2})$ is displayed
as the mass ratio $\alpha$ is increased. For $g_{A}=0$ (Fig.~\ref{cap:GS-alpha},
top), the previously uncorrelated pattern develops into a strongly
localized one ($\alpha=0.02$), where all A atoms are found exclusively
on the \emph{same} site; i.e., they cluster. By contrast, the anti-correlated
{}``insulating'' ground state observed at $g_{A}=0.5$ turns into
a seemingly \emph{uncorrelated} one as the B atom becomes less heavy
(bottom). 

To understand this wealth of phenomena, it would be desirable to extend
the effective Hamiltonian (\ref{eq:H_A}) to higher orders in $\alpha>0$.
This is nontrivial since $H(\alpha)$ in itself does not suggest a
straightforward power series in $\alpha$. Our basic procedure is
sketched in what follows (details will be given elsewhere \citep{pflanzer09a}).
Given the strong localization of the B atom ($N_{B}=1$ without loss
of generality), we introduce center-of-mass and Jacobian coordinates
relative to $x_{B}\approx0$. The resulting Hamiltonian affords an
expansion in powers of $\alpha$, which we can exploit so as to derive
an effective Hamiltonian for the A-atoms by tracing out out B, which
leaves us with $\bar{H}_{\mathrm{A}}(\alpha)=\bar{H}_{\mathrm{A}}^{(0)}+\alpha g_{\mathrm{AB}}\frac{1}{2}\sum_{k\neq l}\left[(x_{A,k}-x_{A,l})n_{B}^{\prime}(x_{A,l})-n_{B}(x_{A,l})\right]+O(\alpha^{2})$.
The first term recovers the initial infinite-mass approximation; the
first-order correction in turn may be understood as an additional
external potential $\delta U_{A}(x)\equiv-\alpha g_{AB}\frac{N_{A}-1}{2}\left[xn'_{B}(x)+n_{B}(x)\right]$,
plus an induced (nonlocal) interaction between two A atoms: $\delta V_{A}(x_{1},x_{2})\equiv\alpha g_{AB}\left[x_{1}n_{B}^{\prime}(x_{2})+x_{2}n_{B}^{\prime}(x_{1})\right]/2$,
which adds to the local one $V_{A}(x_{1},x_{2})=g_{A}\delta(x_{1}-x_{2})$.

Given the effective multi-well (here: two sites) geometry experienced
by the A atoms, it is tempting to explain some qualitative features
from the perspective of simplified lattice models. Since, for $\alpha\to0$,
the single-particle physics of species A is governed by  $\bar{h}_{A}\equiv\frac{1}{2}p^{2}+\frac{1}{2}x^{2}+g_{AB}n_{B}(x)$,
it is natural to use its Bloch-type eigenstates (which for a double
well are delocalized anti-/symmetric  functions that come in bands
of doublets) as a basis for the many-body Hamiltonian. In the spirit
of the (bosonic) Hubbard model \citep{jaksch98}, it is actually preferable
to introduce Wannier-type functions $w_{s}^{(\beta)}$ localized on
the left (right) site $s$ in band $\beta=0,1,\dots$ Expanding the
many-body Hamiltonian $\bar{H}_{\mathrm{A}}(\alpha)$ in terms of
these Wannier functions and---in the case of sufficiently small interaction
energies and deep wells---retaining only lowest-band and on-site terms
yields an effective Bose-Hubbard model\begin{equation}
\bar{H}_{\mathrm{A}}^{(\mathrm{BH})}=-J\sum_{\langle s,s'\rangle}\hat{a}_{s}^{\dagger}\hat{a}_{s'}+\frac{u}{2}\sum_{s}\hat{n}_{s}(\hat{n}_{s}-1).\label{eq:BHM}\end{equation}
Here $-J=\langle w_{s}^{(0)}|\bar{h}_{A}+\delta U_{A}(x)|w_{s'}^{(0)}\rangle$
is the renormalized tunnel coupling of the lowest band, and $u=\langle w_{s}^{(0)\otimes2}|V_{A}+\delta V_{A}|w_{s}^{(0)\otimes2}\rangle=u_{0}+\delta u$
denotes the on-site interaction. While the correction to $J$ tends
to be small, the on-site renormalization due to $\delta V_{A}$ can
have a huge effect if it is on the same order as the $\alpha=0$ term.
A closer analysis \citep{pflanzer09a} reveals that $\delta u<0$
always, signifying an \emph{attractive} induced interaction. There
is an intuitive way of picturing this self-interaction: If an A atom
sits in one well, then it will repel the barrier; thus the well becomes
more spacious, making it energetically favorable to accommodate yet
another A atom. 

Taking the induced attraction at face value, this casts a light on
our results above. For $g_{A}=0$ (Fig.~ \ref{cap:GS-alpha}, top),
$u=\delta u<0$ for a finite mass of B. This makes it plausible that
the A atoms tend to cluster on \emph{same} sites at $\alpha\sim0.02$,
although being really noninteracting. Likewise, for small repulsion
as in Fig.~\ref{cap:GS-alpha} (bottom; $g_{A}=0.5$), both terms
may even cancel, $u_{0}+\delta u\approx0$ -- this helps explain the
seemingly uncorrelated pattern as $\alpha$ is increased.

Although giving the right trend for weak interactions between the
A atoms, the validity of (\ref{eq:BHM}) is actually much more limited
than that of the effective Hamiltonian $\bar{H}_{\mathrm{A}}(\alpha)$.
In particular, the Bose-Hubbard model breaks down for strong interactions
comparable to the band gap. In that regime, rich multi-band effects
can be found for stronger intra-species correlations, which we will
discuss below in the context of the atoms' quantum dynamics.

\paragraph*{Inter-species tunnel dynamics.---}

We have so far investigated the equilibrium situation of species A
in the presence of an effective {}``barrier'' composed of a second,
localized component. It would be thrilling to learn how this affects
the quantum dynamics -- for instance, can the light atoms tunnel through
the heavy ones, and how does the barrier atoms' motion influence this
inter-species tunnel dynamics?

\begin{figure}
\includegraphics[width=0.28\textwidth,keepaspectratio,angle=-90]{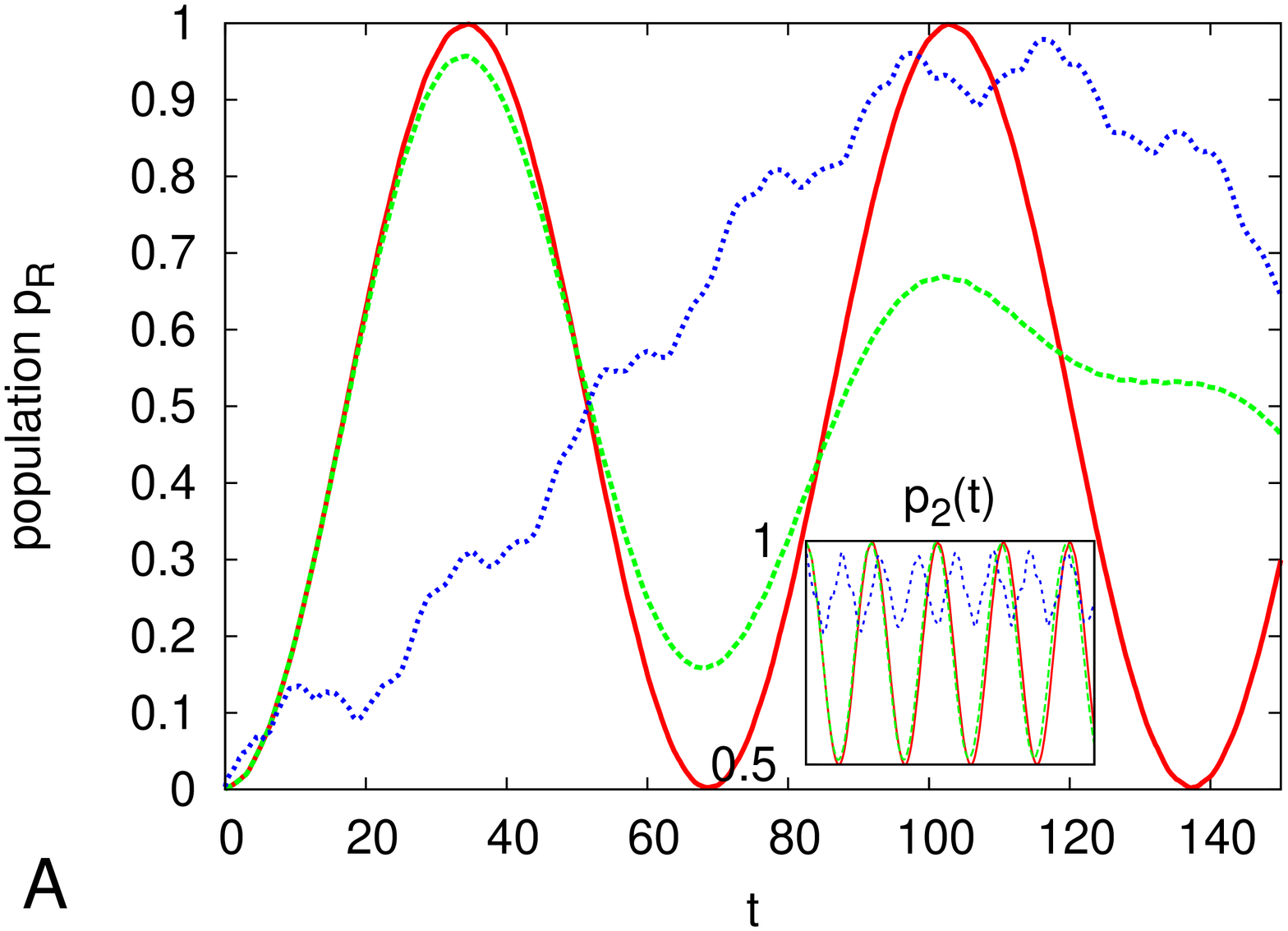}

\includegraphics[width=0.28\textwidth,keepaspectratio,angle=-90]{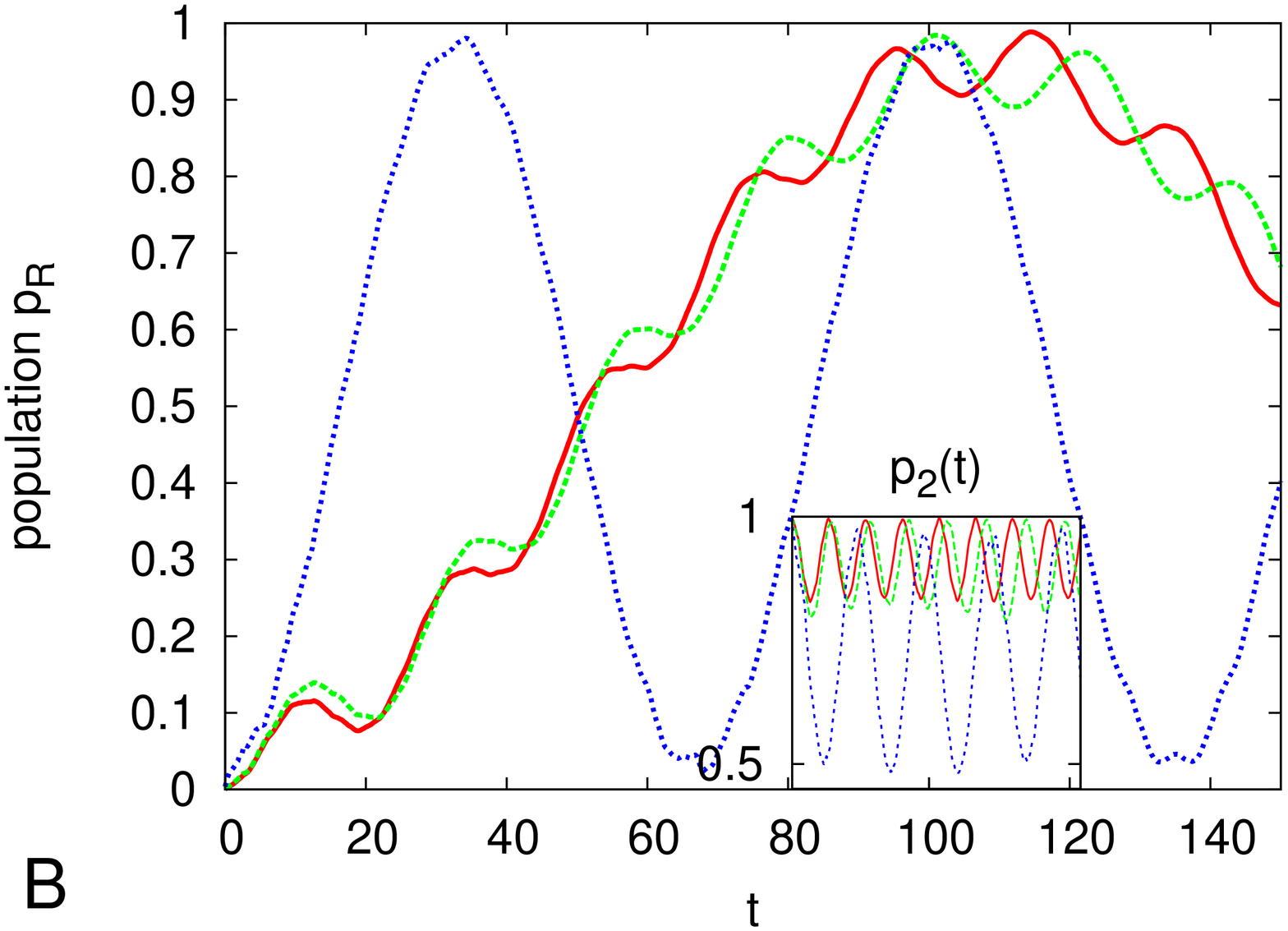}

\includegraphics[width=0.28\textwidth,keepaspectratio,angle=-90]{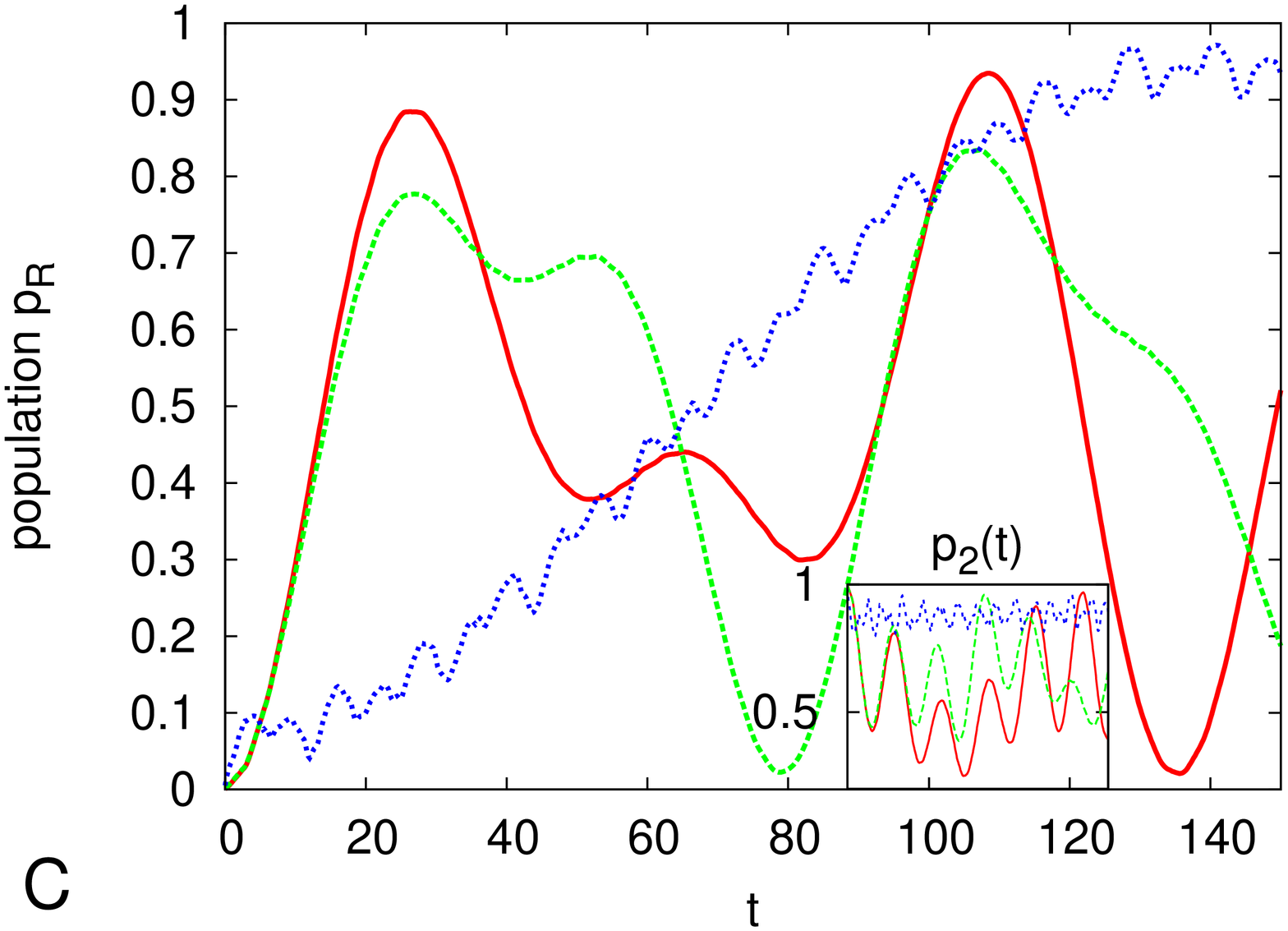}

\caption{(color online) Inter-species tunneling dynamics of $N_{A}=2$ bosons
through a $B$ atom: Relative $A$ population of the right-hand side
over time, $p_{\mathrm{R}}(t)$, for different mass ratios, $\alpha=0.001$
(\textbf{\textcolor{red}{---}}), $\alpha=0.01$ (\textbf{\textcolor{green}{-
- -}}), and $\alpha=0.12$ (\textcolor{blue}{${\color{blue}\boldsymbol{\cdots}}$}).
(a) $g_{A}=0$, (b) $g_{A}=0.5$, (c) $g_{A}=25$; all $g_{AB}=8$.
\emph{Insets}: Probability $p_{2}(t)$ of finding two $A$ atoms on
the same side. \label{cap:tunnel}}

\end{figure}

To answer this question, let us investigate the time evolution of
the A bosons loaded initially in, say, the left-hand side of the trap,
with the barrier atoms (B) tightly centered. This can be done, e.g.,
by displacing the trap center of $U_{A}(x)$ or by blocking the other
half with a laser beam. Upon release, the A atoms may tunnel through
species B, this way moving the barrier atoms, which in turn modifies
the effective potential. To monitor the dynamics, we have recorded
the percentage of A-atoms on the right, $p_{\mathrm{R}}(t)=\int_{0}^{\infty}\rho_{A}(x;t)dx$,
in dependence of $\alpha$. 

Figure~\ref{cap:tunnel}(a) displays this population dynamics for
$N_{A}=2$ \emph{noninteracting} A bosons ($g_{A}=0$). For a static
double well, we simply expect a Rabi-type oscillation of the population
between left and right \citep{zoellner07c}. Indeed, for $\alpha=0.001$,
this is what we find. Increasing the mass ratio, the clear sine mode
gives way to a more complex, two-mode oscillation ($\alpha=0.01$),
until for much larger values ($\alpha=0.12$) the tunneling slows
down drastically, with only a tiny faster modulation on top. This
is reminiscent of second-order tunneling well known from repulsive
atom pairs \citep{foelling07,zoellner07c}. In fact, the effective
Bose-Hubbard model (\ref{eq:BHM}) suggests that this corresponds
to \emph{attractively} bound pairs in a double well, where single-atom
Rabi tunneling is highly suppressed \citep{piil07}. This line of
reasoning is supported by the \emph{pair} (or \emph{same-site}) \emph{probability}
$p_{2}(t)=\int_{\{x_{1}\cdot x_{2}\ge0\}}\rho_{A}^{(2)}(x_{1},x_{2};t)dx_{1}dx_{2}$,
measuring how likely it is to find two A-atoms on the same site \citep{zoellner07c}:
This is plotted in the inset of Fig.~\ref{cap:tunnel}(a), indicating
that the atom pair becomes more and more stable (or $p_{2}$ no longer
drops far below $1$) as $\alpha$ is increased.

This contrasts with the case of $g_{A}=0.5$, shown in Fig.~\ref{cap:tunnel}(b).
Near the static limit ($\alpha\le0.01$), one recovers the situation
of repulsively bound pairs \citep{foelling07,zoellner07c}, which
tunnel at a period $T/2\pi\sim u/4J^{2}$ long compared with the Rabi
oscillations $1/2J$, and which are stable in time (inset). Allowing
for a finite mass of B, this pair breaks up, and the dynamics starts
to resemble sinusoidal Rabi oscillations for $\alpha=0.12$. This
becomes even more conclusive from the angle of the pair probability
(inset), whose minimum value now significantly deviates from unity.
Qualitatively, this feature is captured by the effective Bose-Hubbard
model (\ref{eq:BHM}): At large enough $\alpha$, the attractive on-site
interaction $\delta u<0$ tends to cancel $u_{0}$, which is in agreement
with the Rabi-like oscillations observed here.

What happens for increasing intra-species repulsion, in particular
upon approaching the fermionization limit, e.g., at $g_{A}=25$ {[}Fig.~\ref{cap:tunnel}(c){]}?
In the case of a quasi-static effective barrier ($\alpha=0.001$),
the dynamics resembles that known from fragmented-pair tunneling in
a double well \citep{zoellner07c}. By the Bose-Fermi duality, this
may be regarded as two noninteracting fermions Rabi-tunneling independently
in the lowest $N_{A}=2$ bands $\beta=0,1$: $p_{\mathrm{R}}(t)=\sum_{\beta}\sin^{2}\left(J^{(\beta)}t\right)/N_{A}$.
With increasing $\alpha$, from our previous discussion we expect
an effective attraction between these two fermions. In the intermediate
regime ($\alpha=0.01$), this bears little effect and only leads to
a renormalization of the two {}``Rabi'' frequencies $J^{(\beta)}$.
However, for mass ratios as large as $\alpha=0.12$, the picture changes
qualitatively: The two atoms tunnel only on a time scale about four
times longer than the Rabi oscillations. This pattern closely resembles
that of correlated pair tunneling. Indeed, a look into the two-body
correlations {[}Fig.~\ref{cap:tunnel}(c), inset{]} reveals that
the pair probability $p_{2}$ stays remarkably close to unity, in
marked contrast with the conventional fragmented-atom pair. In this
light, it is enticing to think of this as the tunneling of an attractively
\emph{bound pair} \emph{of identical fermions}.

Let us account for this in a simplified model. Modifying the derivation
of  (\ref{eq:BHM}) for the case of noninteracting \emph{fermions},
we find an effective multi-band Hubbard model \citep{pflanzer09a}.
For the special case of $N_{A}=2$ {}``fermions'', this takes the
form\[
\bar{H}_{\mathrm{A}}^{(\mathrm{FH})}=-\sum_{\langle s,s'\rangle,\beta}J^{(\beta)}\hat{f}_{s}^{(\beta)\dagger}\hat{f}_{s'}^{(\beta)}+\delta u\sum_{s}\hat{n}_{s}^{(0)}\hat{n}_{s}^{(1)},\]
where $N_{A}$ bands $\beta=0,1$ contribute; note that we have discarded
the on-site term $\sum_{s,\beta}\epsilon^{(\beta)}\hat{n}_{s}^{(\beta)}$
constant in our setup. The induced on-site interaction turns out to
be negative, $\delta u<0$. For larger $\alpha$, the induced interaction
$|\delta u|\gg J^{(\beta)}$ shifts single-atom tunneling far off
resonance. It is in this limit that we can understand the inter-species
tunneling as that of an attractively bound fermion(ized) pair at a
fairly large tunnel period $T/2\pi\sim|\delta u|/4J^{(0)}J^{(1)}$.
From this viewpoint, the two quasi-bound fermions reside in different
bands, i.e, pertain to different pseudo-spins and thus are not constrained
by Pauli's principle. This picture extends to arbitrary numbers of
fermions $N_{\mathrm{A}}>2$, which tend to tunnel in dynamically
bound $N_{\mathrm{A}}$-atom clusters.

In conclusion, we have investigated the tunneling of bosonic atoms
through a second, localized, species. Remarkably, this can be well
understood as an induced \emph{attraction} between the mobile bosons.
For the ground state, it may lead to strong intra-species correlations
even in the absence of intra-species interactions.  The inter-species
dynamics, among other things, features tunneling of essentially an
attractively bound fermionized pair. Extending the discussion of tunneling
to nonclassical potential barriers opens up intriguing perspectives,
such as creating an effective lattice made of localized atoms. This
would allow for the study of disorder beyond the {}``quenched''
static limit, to give but one example.

Financial support from Landesstiftung Baden-Württemberg and from Leopoldina
(S.Z.) is gratefully acknowledged. The authors thank H.-D.~Meyer,
S.~Jochim, T.~Gasenzer, and A.~Jackson for valuable discussions.

\bibliographystyle{/home/nbia/zoellner/bin/prsty}
\bibliography{/home/nbia/zoellner/bib/phd,/home/nbia/zoellner/bib/mctdh}

\end{document}